\documentclass{ws-p8-50x6-00}

\newcommand{\ncm}{\newcommand}
\newcommand{\rencm}{\renewcommand}
 
\def\a{\alpha}

\def\f{\varphi}    

\def\l{\lambda}
\def\m{\mu}

\def\o{\omega}
\def\p{\pi}       
\def\r{\rho}      

\def\D{\Delta}

\def\fh{\hat{\f}}

\ncm{\dsp}{\displaystyle}
\ncm{\nn}{\nonumber}
\ncm{\nnn}{\nonumber\linebreak[4]}
\ncm{\nit}{\noindent}
\ncm{\del}{\partial}
\ncm{\av}[1]{\mbox{$\langle #1 \rangle$}}
\ncm{\avc}[1]{\mbox{$\langle #1 \rangle_{\psi}$}}
\ncm{\half}{\mbox{{\small $\frac{1}{2}$}} }
\ncm{\quart}{\mbox{{\small $\frac{1}{4}$}} }
\ncm{\tq}{\mbox{{\small $\frac{3}{4}$}} }
\ncm{\third}{\mbox{{\small $\frac{1}{3}$}} }
\ncm{\sixth}{\mbox{{\small $\frac{1}{6}$}} }
\ncm{\eigth}{\mbox{{\small $\frac{1}{8}$}} }
\ncm{\thrhalf}{\mbox{{\small $\frac{3}{2}$}} }
\ncm{\thrfor}{\mbox{{\small $\frac{3}{4}$}} }
\ncm{\twothi}{\mbox{{\small $\frac{2}{3}$}} }
\ncm{\fivtwo}{\mbox{{\small $\frac{5}{2}$}} }

\ncm{\dxx}{\mbox{$\partial_{x}^2$}}
\ncm{\dx}{\mbox{$\partial_{x}$}}
\ncm{\dt}{\mbox{$\partial_{t}$}}
\ncm{\dtt}{\mbox{$\partial_{t}^2$}}
\ncm{\un}{1\!\!1}
\ncm{\RE}{\mbox{Re}}
\ncm{\IM}{\mbox{Im}}
\ncm{\Tr}{\mbox{tr}\,}
\ncm{\diag}{\mbox{diag}\,}
\ncm{\Det}{\mbox{Det}\,}
\ncm{\Log}{\mbox{Log}\,}
\ncm{\raw}{\rightarrow}
\ncm{\law}{\leftarrow}
\ncm{\dg}{\dagger}
\ncm{\pr}{\prime}
\ncm{\ha}{\hat{a}}
\ncm{\hP}{\hat{P}}

\ncm{\aplt}{ \mbox{}_{\textstyle \sim}^{\textstyle < }     }
\ncm{\apgt}{ \mbox{}_{\textstyle \sim}^{\textstyle > }     }
\ncm{\Oa}{\mbox{$\mbox{O}(a)$}}
\ncm{\Sp}{\mbox{\hspace{1.0cm}}}
\ncm{\capit}[1]{\caption{\it #1}}

\def\be{\begin{equation}}
\def\ee{\end{equation}}
\def\bea{\begin{eqnarray}}
\def\eea{\end{eqnarray}}
\def\bi{\begin{itemize} \itemsep = 0.01\itemsep  }
\def\bii{\begin{itemize} \itemsep = 0.01\itemsep }
\def\ei{\end{itemize}}
\def\bc{\begin{center}}
\def\ec{\end{center}}
\def\bs{\begin{slide}}
\def\es{\end{slide}}

\def\beac{\begin{eqnarray} \color [rgb] {0,0,1} }
\def\eeac{\end{eqnarray} }

\ncm{\shead}[1]{\bc{ \Large \color [rgb]{1.0,.0,.1} #1 \normalcolor} \ec}
\ncm{\ssubh}[1]{{\large \color [rgb]{1.0,.0,.1} #1 \normalcolor}}

\rencm{\thefootnote}{\mbox{\protect{$\fnsymbol{footnote}$}} }

\ncm{\front}[5]   
{
   \begin{titlepage}
      \noindent {#1} \hfill {#2}\\
      \begin{center}
         \vspace{1.5\baselineskip}
         {\Large\bf  #3  } \\
         \vspace{2\baselineskip}
         \vspace{1.5\baselineskip}
          #4\\
         \vspace{1.5\baselineskip}
   
         Institute of Theoretical Physics, \\
         Valckenierstraat 65, 1018 XE Amsterdam,
         The~Netherlands.
    
      \end{center}
      \vfill
      {\bf Abstract}\\
       #5
   \end{titlepage} 
}

\ncm{\frontslide}[4]
{
   \begin{titlepage}
      \noindent {#1} \hfill {#2}\\
      \begin{center}
         \vspace{1.5\baselineskip}
         {\Large\bf  #3  } \\
         \vspace{2\baselineskip}
         \vspace{1.5\baselineskip}
          #4\\
         \vspace{1.5\baselineskip}
   
         Institute of Theoretical Physics, \\
         Valckenierstraat 65, 1018 XE Amsterdam,
         The~Netherlands.
    
      \end{center}
   \end{titlepage} 
}

\begin{document}

\title{Scalar field dynamics: classical, quantum and in between\footnote{ 
\noindent   \lowercase{\uppercase{P}resented by \uppercase{J}.~\uppercase{V}ink,
 at ``\uppercase{S}trong and 
          \uppercase{E}lectroweak \uppercase{M}atter" 
          (\uppercase{SEWM}2000),
 \uppercase{M}arseille, \uppercase{F}rance, \uppercase{J}une 14-17, 2000. }} 
}

\author{ M.~Sall\'e, J.~Smit and J.C.~Vink }

\address{Institute for Theoretical Physics, Valckenierstraat 65, 
1018 XE Amsterdam,\\ The Netherlands
}


\maketitle

\abstracts{
Using a Hartree ensemble approximation, we investigate the dynamics of 
the $\f^4$ model in $1+1$ dimensions. We find that the fields initially 
thermalize with a Bose-Einstein distribution for the fields. Gradually, 
however, the distribution changes towards classical equipartition. 
Using suitable initial conditions quantum thermalization is achieved much 
faster than the onset of this undesirable equipartition. We also show how 
the numerical efficiency of our method can be significantly improved. }

\section{Inhomogeneous Hartree Dynamics}

In many areas of high-energy physics, e.g. heavy ion collisions and early
universe physics,  a non-perturbative understanding of quantum field 
dynamics is required. Computer simulations could fulfill this need,
but as is well-known the problem is very difficult.
Hence one resorts to approximations, such as
classical dynamics\cite{GrRu88} (for recent work see\cite{Aart00}),
large $n$ or Hartree (see e.g.\ \cite{CoHa98}). 
Here we introduce and apply a different type
of Hartree or gaussian approximation than used previously, in which 
we use an ensemble of gaussian wavefunctions to compute Green functions.
We can only sketch this method here, a detailed 
presentation will appear elsewhere.

We use the lattice $\f^4$ model in $1+1$ dimensions as a test model, which has
the following Heisenberg operator equations,
\be
\dot{\hat{\f}} = \hat{ \p}, \;\;\;
\dot{\hat{\p}} = (\D - \m^2)\hat{\f} - \l \hat{\f}^3, \label{eq:heis}
\ee
with $\D$ the lattice laplacian, $\m$ the bare mass and $\l$ the coupling 
constant.
Rather than solving the operator equation (\ref{eq:heis}) in all detail,
one may focus on the Green functions.
The Hartree approximation assumes that the density matrix
used to compute these Green functions is of gaussian form, such that
all information is contained in the one- and two-point functions.
The one-point function is the mean field and the connected two-point function
is conveniently expanded in terms of mode-functions,
\be
 \av{ \fh_x } = \f_x,\;\;
 \av{ \fh_x \fh_y }_{\rm conn} =  \sum_\a f_x^\a f_y^{\a *}.
\ee
We have restricted ourselves to {\em pure state} gaussian density matrices here.
The Heisenberg equations now provide self-consistent
equations for the mean field $\f$ and the mode functions $f^\a$,
\bea
   \ddot{\f}_x & = & \D\f_x - [ \m^2 + \l (\f_x^2 +
             3\sum_{\a} f_x^{\a} f_x^{\a *}) ]\f_x \nonumber \\
   \ddot{f}_x^{\a} & = & \D f_x^\a - [ \m^2 + \l (3\f_x^2 +
      3\sum_{\a} f_x^{\a} f_x^{\a *}) ] f_x^\a  \label{eq:eom}
\eea

In order to simulate a more general density matrix than the 
gaussian ones required in the Hartree approximation, we average
over a suitable ensemble of Hartree realizations, by specifying different
initial conditions and/or coarsening in time.
In this way we may compute Green functions with a non-gaussian density matrix 
$\hat\r = \sum_i p_i \hat\r_i^G$, as
\be
 \av{ \hat\f_x \hat\f_y }_{\rm conn} =
   \sum_i p_i [  \av{ \hat\f_x \hat\f_y }_{\rm conn}^{(i)}
                      + \f_x^{(i)} \f_y^{(i)}] - ({\rm \scriptsize disconn.})
\label{eq:avgx}
\ee
Note that the individual Green functions labeled by $i$ are 
still computed with pure 
gaussian states $\hat\r^G_i$, as is appropriate in the Hartree approximation.

The field operator in the Hartree approximation may be written as
\be
\fh_x = \f_x + \sum_\a[ f^\a_x \hat{a}_\a + f^{* \a}_x \hat{a}^{\dagger}_\a],
\ee
with $\hat{a}_\a$ and $\hat{a}^{\dagger}_\a$ time-independent creation and
annihilation operators.
This suggests that the mode functions represent the (quantum) particles in
the model.
It should be stressed that in general $\f_x$ is {\it inhomogeneous} in space.
We also note that the equations (\ref{eq:eom}) 
can in fact be derived from a hamiltonian.
Since the equations are also strongly non-linear, this suggests that the
system will evolve to an equilibrium distribution with equipartition of energy,
as in classical statistical physics.

\section{Observables}
To assess the viability of the Hartree ensemble approximation, we
solve the equations (\ref{eq:eom}) starting from a number of initial conditions. 
With Hartree dynamics we expect to go beyond classical dynamics, because
the width of the (gaussian) wavefunction, represented by the mode functions,
should capture important
quantum effects. Of course we cannot expect to capture everything, e.g.
tunneling is beyond the scope of the gaussian approximation. 

Similarly to using classical dynamics, we expect that
after coarse-graining in space and time and averaging over initial
conditions, we may compute the
particle number densities and energies from the (Fourier transform of the)
connected two-point functions of $\f$ and $\p$,
\be
\frac{1}{N}\sum_{xz}e^{-ikz} \av{\overline{ \hat\f_{x+z} \hat\f_x} }_{\rm conn}
    = \frac{n_k + \half}{\o_k},\;\;\;
\frac{1}{N}\sum_{xz}e^{-ikz} \av{ \overline{ \hat\p_{x+z} \hat\p_x} }_{\rm conn}
    =  (n_k + \half) \o_k.
\label{eq:nando}
\ee
The over-bar indicates averaging over some time-interval and initial conditions
as in eq. (\ref{eq:avgx}),
$\o_k$ is the energy, $n_k$ the (number) density of particles with
momentum $k$ and $N$ the number of lattice sites. 

For weak couplings, such as we will use
in our numerical work, the particle densities should have a Bose-Einstein (BE)
distribution,
\be
  n_k = 1/(e^{\o_k/T} - 1),\;\;\; \o_k = (m^2(T) + k^2)^{\half},
\label{eq:freeform}
\ee
with $T$ the temperature and $m(T)$ an effective  finite temperature mass.

\begin{figure}[t]
\hspace{0.5cm}
\scalebox{0.35}{ \rotatebox{270}{\includegraphics{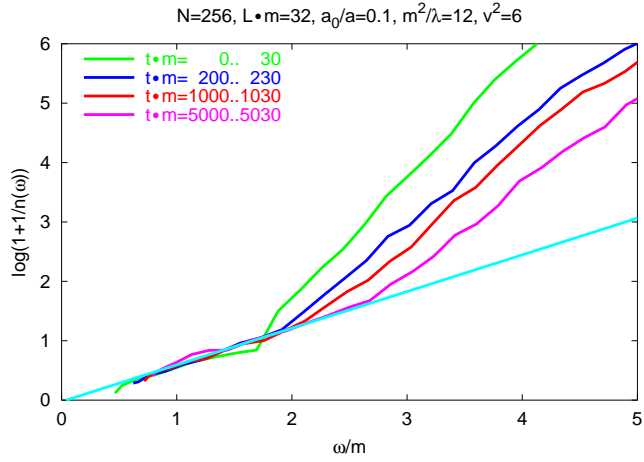} } }
\caption{Particle number densities as a function of $\o$ at various 
times. The straight line gives a BE distribution with temperature
$T/m = 1.7$. (The lattice volume $Lm=32$, coupling $\l/m^2=0.083$ 
and inverse lattice spacing $1/am = 8$; $m \equiv m(T=0)$).
\vspace{-0.7cm}
\label{fig:pa}}
\end{figure}

\section{Numerical results}
First we use initial conditions which correspond to fields far out of
equilibrium: gaussians with mean fields that consist of just
a few low momenta modes,
\be
\f_x = v,\; \p_x = A \sum_j^{j_{\rm max}}   \cos(k_j x + \a_j).
\ee
The $\a_j$ are random phases and $A$ is a suitable amplitude.
Initially the mode functions are plane waves, $e^{ik x}/\sqrt{2\o L}$,
i.e. there are no quantum particles and all energy
resides in the mean field.

\begin{figure}[t]
\hspace{-0.6cm}
\scalebox{0.74}[1.11]{ \includegraphics[clip]{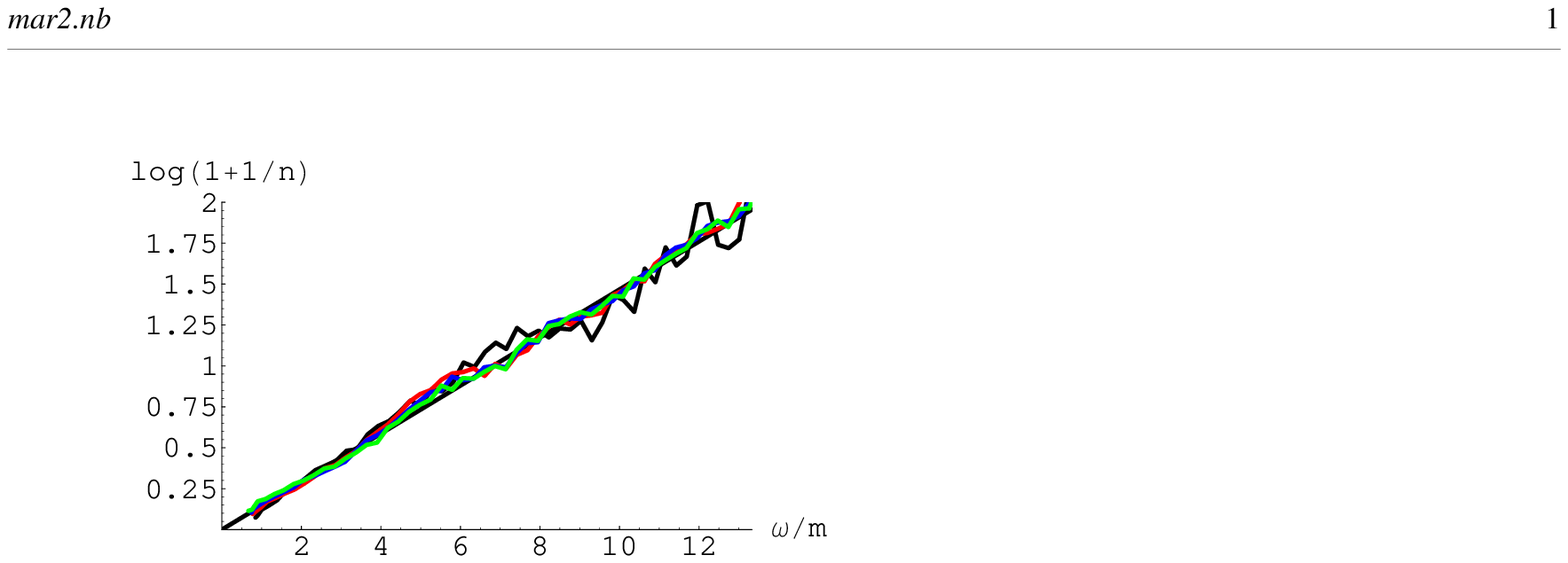} } \hspace{-0.7cm}
\scalebox{0.74}[1.06]{ \includegraphics[clip]{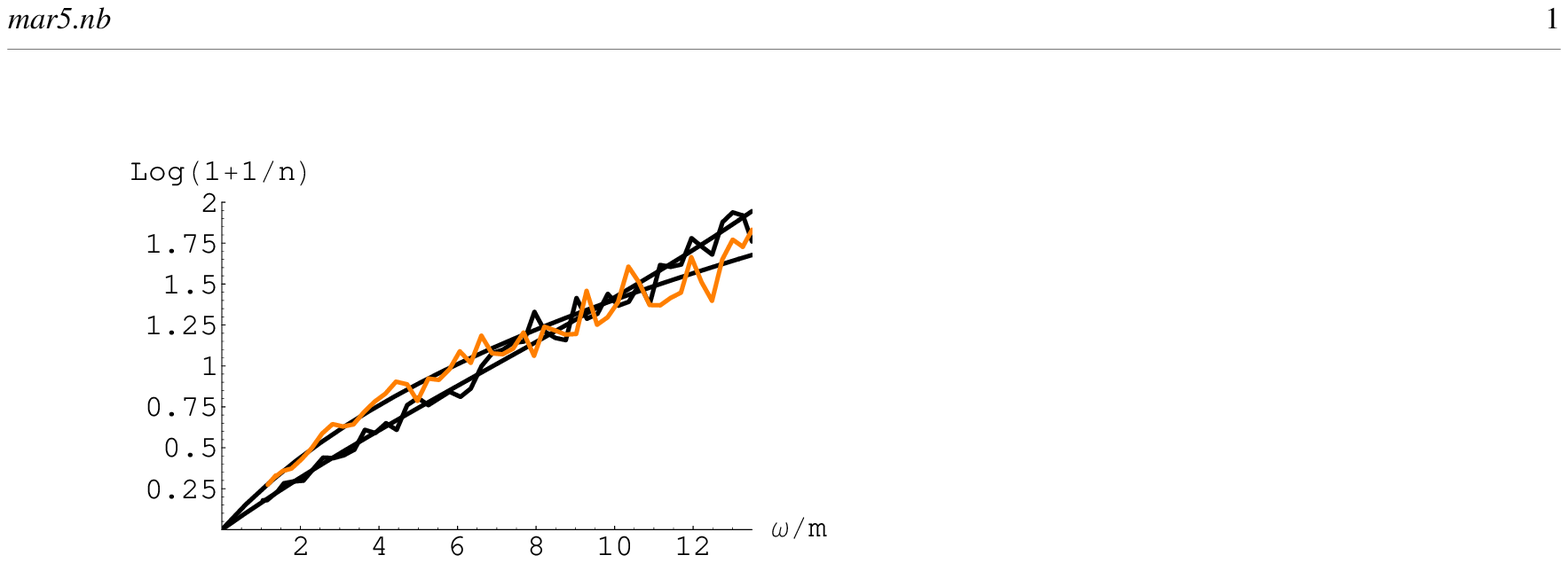}} 
\vspace{5.0cm}
\caption{Left: Particle number densities as a function of $\o$ at times 
$tm=100$, $1400$, $2800$ and $4200$ (bottom up),
starting from BE type initial conditions for the mean field.
The straight line through the origin is a BE distribution with
temperature $T/m=6.7$.
The right figure uses classical dynamics and shows data at $tm=100$
and $tm=4200$.  ($Lm=23$, $\l/m^2=0.083$ and $1/am = 22$).
\label{fig:be}}
\vspace{-0.7cm}
\end{figure}

The results in Fig.~\ref{fig:pa} show that very fast, $tm\aplt 200$, 
a BE distribution
is established for particles with low momenta. Slowly this thermalization
progresses to particles with higher momenta, while the temperature 
$T \approx 1.7m$ remains roughly constant. Such a thermalization does
{\em not} happen when using {\em homogeneous} Hartree dynamics ($\f$
constant in space). The difference may be understood, because
particles in our method can scatter off the 
inhomogeneities in the mean field.

Next we want to speed-up the thermalization of the high momentum modes.
Therefore we use different initial conditions in which the energy
is distributed more realistically over the Fourier modes of the mean
field. The modes are the same as before but mean fields are drawn from
an ensemble with a BE-like probability distribution,
\be
 P(\f_k, \p_k) \propto \mbox{exp}
          [ -(e^{\o_k/T_0}-1)( \p_k^2  + \o_k^2 \f_k^2)/2\o_k ]
\ee
Now we find density distributions as shown in Fig.~\ref{fig:be} (left).
Already after a short time, $tm\approx 100$ the particles have acquired a
BE distribution up to large momenta $\o/m\approx 12$.
Note that even with these BE-type initial conditions, the fields initially are
still out of equilibrium: energy is initially carried by the mean
field only but is quickly, within a time span of $tm\approx 200$,
redistributed over the modes.

\begin{figure}[t]
\hspace{0.8cm}
\scalebox{1.05}[1.00]{ \includegraphics[clip]{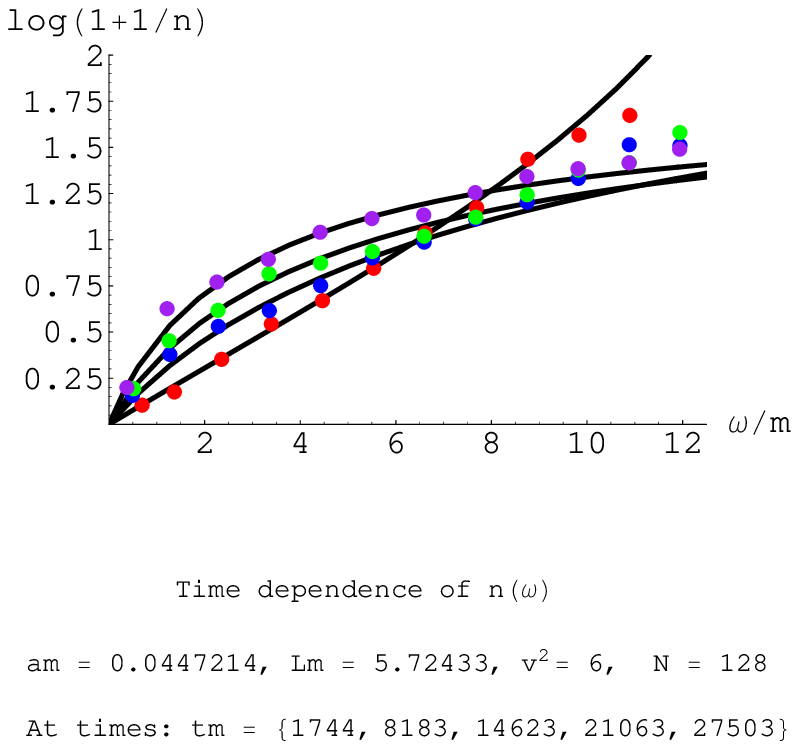} } 
\vspace{4.9cm}
\caption{Particle number densities as a function of $\o$ at times 
$tm\approx 1700$ (straight line), and $8000$, $15000$ and $27000$ (increasingly
curved lines). 
The drawn lines are fits with Ansatz $n = c_0 + c_1/\o$.
 ($Lm=5.7$, $\l/m^2=0.083$ and $1/am = 22$).
\label{fig:long}}
\vspace{-0.5cm}
\end{figure}

Since the Hartree dynamics follows from a hamiltonian, one might
expect that eventually equipartition sets in.
To investigate this, we have run a simulation
for long times, with BE-type initial conditions. The results for
several times are shown in Fig.~\ref{fig:long}. The (curved) lines here
are fits to the data with Ansatz $n = c_0 + c_1/\o$. 
For times $tm\aplt 2000$, the data show a BE-distribution, which was
established within $tm\aplt 200$. At later times  $tm \gg 2000$ one 
recognizes classical equipartition with temperature $T=c_1$ and $c_0 \approx 0$.
Since the energy is gradually distributed over particles with increasingly
larger momentum range, the temperature slowly decreases.

To see the difference between our Hartree and purely classical
dynamics, we have repeated the simulation of Fig.~\ref{fig:be} (left), now
using the classical e.o.m. for $\f$. As can be seen in Fig.~\ref{fig:be} (right)
the initial BE distribution with $T=6.7m$ evolves to equipartition much faster.
Already at $tm \approx 4200$ we find that the distribution is well
represented by the classical form $n = 3.7m/\o$ (the curved
line in Fig.~\ref{fig:be} (right)).

\begin{figure}[t]
\scalebox{0.72}[0.72]{ \includegraphics[clip]{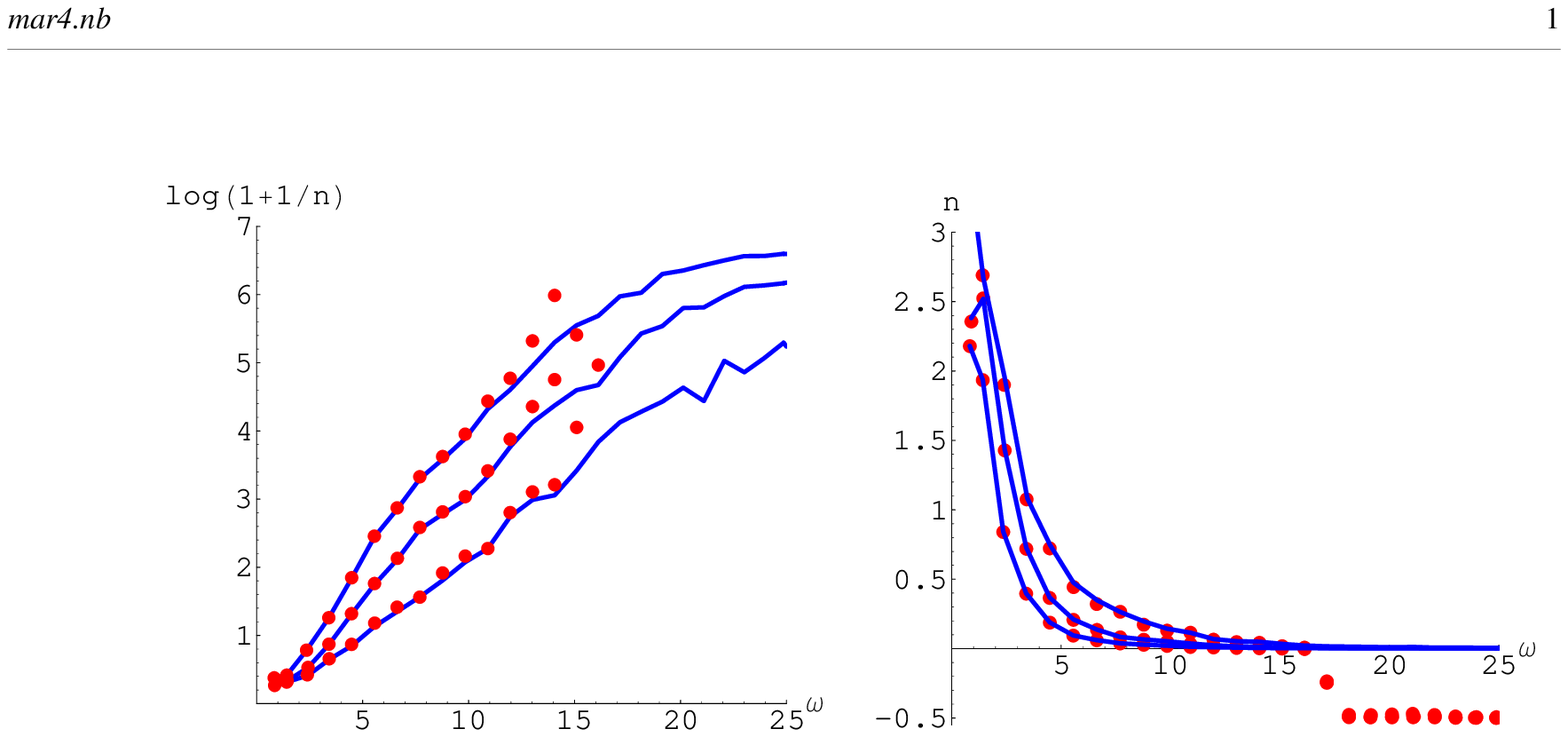} }
\vspace{4.2cm}
\caption{ Hartree dynamics with all mode functions (drawn lines)
and with a reduced number (32 out of 128) mode functions (dots).
($Lm=5.7$, $\l/m^2=0.083$ and $1/am = 22$).
\label{fig:comp}}
\vspace{-0.7cm}
\end{figure}

Finally we try to improve the efficiency of our method. Since we have
to solve for $N$ mode functions, on a lattice with $N$ sites, the CPU
time for one time-step grows $\propto N^2$. However, mode functions 
corresponding
to particles with momenta much larger than $T$ should be irrelevant,
since these particle densities are exponentially suppressed. This suggests
that we can discard such modes. This is tested in Fig.~\ref{fig:comp},
where we compare simulation results using all mode functions with results
obtained using only a quarter of the mode functions. This corresponds with
mode functions with initial plane wave energy $\o \aplt 17m$. 
Clearly the results
for particles with significant densities $n$ are indistinguishable.
For particles with energy larger than $\approx 17m$, there are no longer
mode functions that can provide the vacuum fluctuations and consequently
the particle density defined by (\ref{eq:nando}) drops to $-1/2$.

\section{Conclusion}
We have demonstrated that, using our Hartree ensemble method,
we can simulate quantum thermalization in a simple scalar field model in 
real time.
Only after times much longer than typical equilibration and damping
times,  the approximate nature of the dynamics shows up in deviations
from the BE distribution towards classical equipartition. See the 
contribution of Smit in these proceedings for an
estimate of damping times in our model\cite{Smit00}.
We have furthermore shown that these simulations can be done using a 
limited number of mode functions: only mode functions for particles with 
energies below a few times the temperature are required.

\end{document}